\documentclass[preprintnumbers,floatfix,letterpaper,superscriptaddress,nofootinbib,twocolumn]{revtex4}
\usepackage{graphicx}
\usepackage{microtype}
\usepackage{amsmath}
\usepackage{amssymb}
\usepackage{subfigure}
\usepackage{hyperref}
\usepackage{url}
\usepackage{xcolor}
\usepackage{color}
\usepackage{mathrsfs}
\usepackage{calrsfs}
\usepackage{amsfonts}
\usepackage{latexsym}
\usepackage{ragged2e}
\usepackage{epsfig}
\usepackage{textcomp}
\usepackage{caption}

\DeclareCaptionJustification{justified}{\leftskip=0pt
\rightskip=0pt \parfillskip=0pt plus 1fil}
\definecolor{vividviolet}{rgb}{0.62, 0.0, 1.0}
\definecolor{amaranth}{rgb}{0.9, 0.17, 0.31}
\definecolor{palatinateblue}{rgb}{0.15, 0.23, 0.89}
\definecolor{brightpink}{rgb}{1.0, 0.0, 0.5}
\definecolor{cornflowerblue}{rgb}{0.39, 0.58, 0.93}
\definecolor{deepcarminepink}{rgb}{0.94, 0.19, 0.22}
\definecolor{radicalred}{rgb}{1.0, 0.21, 0.37}
\hypersetup{ linktoc=all,
    colorlinks, linkcolor={palatinateblue},
    citecolor={brightpink}, urlcolor={amaranth}
}

\graphicspath{{Images/}}

\makeatletter
\def\@fnsymbol#1{\ensuremath{\ifcase#1\or \dagger\or \ddagger\or $\textleaf$
\else\@ctrerr\fi}}

\makeatother

\begin{document}

\title{\textbf{\ Black Hole Remnant in Massive Gravity}}
\affiliation{{\footnotesize {Center for Gravitation and Cosmology, College of Physical
Science and Technology,}}\\
Yangzhou University, Yangzhou 225009, China}
\affiliation{{\footnotesize Research Institute for Astronomy and Astrophysics of Maragha
(RIAAM), P.O. Box 55134-441, Maragha, Iran}}
\affiliation{{\footnotesize Physics Department and Biruni Observatory, College of
Sciences, Shiraz University, Shiraz 71454, Iran}}
\affiliation{{\footnotesize {ICRANet, Piazza della Repubblica 10, I-65122 Pescara, Italy}}}
\affiliation{{\footnotesize {School of Physics and Astronomy, Shanghai Jiao Tong
University, Shanghai 200240, China}}}
\affiliation{{\footnotesize {Nordita, KTH Royal Institute of Technology \& Stockholm
University, Roslagstullsbacken 23, SE-106 91 Stockholm, Sweden}}}
\author{B. Eslam Panah}
\email{beslampanah@shirazu.ac.ir}
\affiliation{{\footnotesize Research Institute for Astronomy and Astrophysics of Maragha
(RIAAM), P.O. Box 55134-441, Maragha, Iran}}
\affiliation{{\footnotesize Physics Department and Biruni Observatory, College of
Sciences, Shiraz University, Shiraz 71454, Iran}}
\affiliation{{\footnotesize {ICRANet, Piazza della Repubblica 10, I-65122 Pescara, Italy}}}
\author{S. H. Hendi}
\email{hendi@shirazu.ac.ir}
\affiliation{{\footnotesize Physics Department and Biruni Observatory, College of
Sciences, Shiraz University, Shiraz 71454, Iran}}
\author{Y. C. Ong}
\email{ycong@yzu.edu.cn}
\affiliation{{\footnotesize {Center for Gravitation and Cosmology, College of Physical
Science and Technology,}}\\
Yangzhou University, Yangzhou 225009, China}
\affiliation{{\footnotesize {School of Physics and Astronomy, Shanghai Jiao Tong
University, Shanghai 200240, China}}}
\affiliation{{\footnotesize {Nordita, KTH Royal Institute of Technology \& Stockholm
University, Roslagstullsbacken 23, SE-106 91 Stockholm, Sweden}}}

\begin{abstract}
The possibility of a nonzero graviton mass has been widely pursued in the
literature. In this work we investigate a black hole solution in massive
gravity with a degenerate fiducial metric often used in the literature. We
find that the end state of Hawking evaporation leads to black hole remnant,
which could help to ameliorate the information paradox. We prove that these
remnants only exist in anti-de Sitter spacetime. Nevertheless, we speculate
on their possible relevance to our Universe as dark matter candidate, in
view of the possibility that our Universe could be \emph{inherently} anti-de
Sitter-like, with a transient accelerated expansion phase.
\end{abstract}

\maketitle

%%%%%%%%%%%%%%%%%%%%%%%%%%%%%%%%%%%%%%%%%%%%%%%%%%%%%%%%%%%%%%%%%%%%%%%%%%%%%%%%%%%%%%%%%%%%%%%%%%%%%%%%%%%

%%%%%%%%%%%%%%%%%%%%%%%%%%%%%%%%%%%%%%%%%%%%%%%%%%%%%%%%%%%%%%%%%%%%%%%%%%%%%%%%%%%%%%%%%%%%%%%%%%%%%%%%%%%

\section{Introduction: A Brief Account of Massive Gravity}

\label{s1}

%Ever since astrophysical and cosmological observations suggested that some form of dark matter exist, there have been no shortage for dark matter candidates, though after decades of various experiments, none of them have been found. Although most dark matter candidates stem from particle physics (e.g. supersymmetric particles), there is an intriguing possibility that they could be black hole \emph{remnants} \cite{MacGibbon,Barrow,Carr}, which are the end state of Hawking evaporation from primordial black holes. If so, then there is no need for a new kind of unobserved elementary particle to explain something that apparently populates a quarter of the Universe. The price to pay is that one has to explain how black hole remnants arise, since they are not part of the usual Hawking process.

Black hole remnants are the stable or meta-stable end state of Hawking
evaporation, in the sense that Hawking radiation may stop as the mass of the
black hole reaches the Planck scale, due to new physics of quantum
gravitational nature. They can arise from different theories, or from
various quantum gravity inspired phenomenological models. Properties of
black hole remnants have been studied in the literature \cite{Giddings,
Giddings2}. See \cite{Chen} for a recent review of the subject. In this
paper, our main objective is \emph{to find black hole remnants in the theory
of massive gravity, without additional gauge fields}, which to our knowledge
has not been explored before. Given that massive gravity is nowadays a
popular candidate for modified theory of gravity (despite its various short
comings), this is a topic worth studying -- can black hole remnant arise
simply by endowing graviton with a mass?

We begin with some background on massive gravity for completeness.
Einstein's general relativity can be cast as a theory of massless spin-$2$\
gravitons. Generalizations to massive gravity theories have several
motivations, including an attempt to explain the observed accelerated expansion of the Universe. 
One could also investigate massive
gravity as an extension of general relativity, to see if such a theory is
consistent, or if general relativity is the unique consistent spin-2 theory
of gravitation. Recent observations by LIGO has put a tight bound on
graviton's mass \cite{BoundIII,BoundIV}, but well-known work of
Boulware and Deser \cite{Boulware} showed that a generic extension of the
Fierz-Pauli (FP) theory \cite{FP} to curved backgrounds will give rise to ghost
instabilities, now known as the \textquotedblleft BD-ghost\textquotedblright
. A special generalization to a nonlinear and stable massive gravity has
been introduced by de Rham, Gabadadze and Tolley (dRGT) \cite{dRGT0,dRGT,
Rham}, which was subsequently shown to be ghost-free by Hassan and Rosen
\cite{Hassan,HassanI,HassanII}. As we shall see below, dRGT massive gravity
comes with two metric tensors, one of which is a fixed spacetime background,
as realized by Hassan and Rosen. A natural generalization to having both
gravitons being dynamical was then sought by Hassan and Rosen, later known
as the bimetric or bi-gravity theory \cite{1109.3515}. However, in this
work, we shall focus on the original massive gravity with a fixed background.

In addition to cosmological implications, the nonzero graviton mass allows
one to model field theories with momentum dissipation in holography, without
the need to employ the more traditional lattice method in the anti-de Sitter
bulk \cite{Vegh, 1308.4970}.

We note that dRGT massive gravity suffers from some problems. Firstly, there
is a lack of viable FLRW cosmological solutions \cite{DAmico,1304.0484}.
More accurately, massive gravity does not admit FLRW solutions if a flat
reference metric is assumed. However such solutions can exist with other
choices of the reference metric.\textbf{\ }

There are also fundamental problems related to the well-posedness of the
theory, and \textquotedblleft micro-acausality\textquotedblright\
(arbitrarily small closed causal curve) \cite{1306.5457, 1408.0561,
1410.2289, 1504.02919, 1505.03518} (it is likely that these problems are
avoided in the bimetric generalization, given the recent understanding of
its complicated causal structure \cite{1706.07806}). Nevertheless, there is
still merit in further understanding the various aspects of the theory. For
example, the effects of nonzero graviton mass on the structure of neutron
stars \cite{1701.01039} and white dwarfs \cite{1805.10650} have been studied
recently. The results showed that the maximum mass of these stars can be
about three times the solar mass, i.e. more massive than in general
relativity.

In this work, we will focus on demonstrating that massive gravity admits
black hole remnants. Interestingly, this type of remnants exist in anti-de
Sitter spacetime, but not in de Sitter one. We will discuss its possible
implications for information paradox of black holes. We also speculate on
the relevance of these remnants in our actual Universe as possible dark
matter candidate; although the Universe is currently undergoing accelerated
expansion, it is possible that this is a transient period, and the Universe
is actually inherently anti-de Sitter-like. That is to say, current
observation does not rule out the possibility that our Universe will be
asymptotically anti-de Sitter in the future. The cosmological constant would
eventually dominate the evolution of the Universe, slowing its expansion.
The current phase of accelerating expansion could be the results of other
fields, whose effects might become sub-dominant compared to the cosmological
constant in the far future. A recent study that attempted to address the
current tension regarding values of the Hubble constant measured by low $z$
observations and high $z$ Planck measurement from CMB has also proposed such
a scenario \cite{1808.06623}.

%%%%%%%%%%%%%%%%%%%%%%%%%%%%%%%%%%%%%%%%%%%%%%%%%%%%%%%%%%%%%%%%%%%%%%%%%%%%%%%%%%%%%%%%%

\section{Black Hole Remnants in Massive Gravity}

The action of dRGT massive gravity can be written as Hilbert-Einstein action
with suitable nonlinear interaction terms \cite{dRGT}:
\begin{equation}
I=\frac{1}{16\pi }\int \text{d}^{4}x\sqrt{-g}\left( R+m^{2}\mathcal{U}%
(g,\phi ^{a})\right) ,  \label{action}
\end{equation}%
where $R$ and $\mathcal{U}$ are, respectively, the Ricci scalar and the
effective potential of graviton which modifies the gravitational sector with
a nonzero graviton mass $m$. Note that despite appearance, dRGT gravity
should not be viewed as a \textquotedblleft scalar-tensor\textquotedblright\
theory -- the scalar fields are St\"{u}ckelberg scalars, introduced as a
mean to restore the general covariance of the theory \cite{1105.3735}. Note
that we do not include the cosmological constant a priori in the action,
though of course this can be done just as well as in general relativity. The
reason for omitting the cosmological constant is in view of the original
motivation of the massive gravity theory to explain the accelerating
expansion of the Universe without resorting to a cosmological constant.

The Newton constant is dimensionful, but we will set its \emph{value} as
unity for simplicity. This means that graviton mass $m$ has dimension
inverse length ($c=1=\hbar $), but terms like $M/r\equiv GM/c^{2}r$ are
dimensionless. This follows the convention of, e.g., \cite{Ghosh, Cai,
HendiEP1}. The effective potential $\mathcal{U}$ can be written as
\begin{equation}
\mathcal{U}\left( g,\phi ^{a}\right) =\mathcal{U}_{2}+\alpha _{3}\mathcal{U}%
_{3}+\alpha _{4}\mathcal{U}_{4},  \label{U}
\end{equation}%
in which $\alpha _{3}$ and $\alpha _{4}$ are two dimensionless free
parameters of the theory. The functional form of $\mathcal{U}_{i}$ with
respect to the metric $g$ and scalar field $\phi ^{\alpha }$ are given by
\begin{eqnarray}
\mathcal{U}_{2} &=&\left[ \mathcal{K}\right] ^{2}-\left[ \mathcal{K}^{2}%
\right] ,  \notag \\
\mathcal{U}_{3} &=&\left[ \mathcal{K}\right] ^{3}-3\left[ \mathcal{K}\right] %
\left[ \mathcal{K}^{2}\right] +2\left[ \mathcal{K}^{3}\right] ,  \notag \\
\mathcal{U}_{4} &=&\left[ \mathcal{K}\right] ^{4}-6\left[ \mathcal{K}^{2}%
\right] \left[ \mathcal{K}\right] ^{2}+8\left[ \mathcal{K}^{3}\right] \left[
\mathcal{K}\right] +3\left[ \mathcal{K}^{2}\right] ^{2}  \notag \\
&\left. {}\right. &-6\left[ \mathcal{K}^{4}\right] ,
\end{eqnarray}%
in which
\begin{equation}
\mathcal{K}_{\nu }^{\mu }=\delta _{\nu }^{\mu }-\sqrt{g^{\mu \sigma
}f_{ab}\partial _{\sigma }\phi ^{a}\partial _{\upsilon }\phi ^{b}},
\end{equation}%
where $f_{ab}$ is an appropriate non-dynamical reference metric and the
rectangular bracket denotes the traces, namely $\left[ \mathcal{K}\right] =%
\mathcal{K}_{\mu }^{\mu }$ and $\left[ \mathcal{K}^{n}\right] =\left(
\mathcal{K}^{n}\right) _{\mu }^{\mu }$. In addition, $\phi ^{a}$'s are the St%
\"{u}ckelberg scalars. In the forthcoming discussions, we shall use the
following redefinitions for $\alpha _{3}$ and $\alpha _{4}$, following the
convention of \cite{Ghosh}
\begin{equation}
\alpha _{3}=\frac{\alpha -1}{3},~~~~~\alpha _{4}=\frac{\beta }{4}+\frac{%
1-\alpha }{12},
\end{equation}%
where $\alpha $ and $\beta $ are two arbitrary dimensionless constants.

One finds the following field equation by varying the action with respect to
$g_{\mu \nu }$:
\begin{equation}
G_{\mu \nu }+m^{2}\chi _{\mu \nu }=0,  \label{Field equation}
\end{equation}%
where $G_{\mu \nu }$ is the Einstein tensor and $\chi _{\mu \nu }$ takes the
form
\begin{eqnarray}
\chi _{\mu \nu } &=&\mathcal{K}_{\mu \nu }-\mathcal{K}g_{\mu \nu }-\alpha
\left\{ \mathcal{K}_{\mu \nu }^{2}-\mathcal{KK}_{\mu \nu }+\frac{\mathcal{U}%
_{2}}{2}g_{\mu \nu }\right\}  \notag \\
&&+3\beta ^{2}\left\{ \mathcal{K}_{\mu \nu }^{3}-\mathcal{KK}_{\mu \nu }^{2}+%
\frac{\mathcal{K}_{\mu \nu }}{2}\mathcal{U}_{2}-\frac{1}{6}g_{\mu \nu }%
\mathcal{U}_{3}\right\} .  \notag
\end{eqnarray}

We now consider a $4$-dimensional static, spherically symmetric spacetime
with metric ansatz
\begin{equation}
\text{d}s^{2}=-g(r)\text{d}t^{2}+\frac{\text{d}r^{2}}{g(r)}+r^{2}\left(
\text{d}\theta ^{2}+\sin ^{2}\theta \text{d}\varphi ^{2}\right) .
\label{metric}
\end{equation}

The reference metric essentially plays the role of a Lagrange multiplier to
eliminate the BD ghost, and also different choices of the reference metrics
give different theories. Here we shall follow the same choice for
non-dynamical reference metric in the following form \cite{Vegh,Cai,HendiEP1}%
;
\begin{equation}
f_{ab}=\text{diag}(0,0,c^{2},c^{2}\sin ^{2}\theta ),  \label{reference}
\end{equation}%
where $c$ is a positive constant with dimension of length. We should
emphasize that the property of massive gravity is such that the choice of
reference metric \emph{does} affect what kind of solutions are allowed, so
this black hole solution depends on the choice made above. (For detailed
study on black hole solutions in dRGT theory, see \cite{resultVII})
Admittedly, the proof of ghost-freeness of dRGT theory \cite{HassanI,
HassanII} assumes that the reference metric is invertible, so for degenerate
metric (i.e. its rank is smaller than its dimension) like Eq. (\ref%
{reference}) one has to analyze the BD ghost separately \cite{Vegh,
1308.4970}. It was shown that in \cite{1510.03204}, the aforementioned
non-dynamical reference metric in Eq.(\ref{reference}), does indeed give
rise to ghost-freeness. However, since the existence of BD ghost depends not
only on the background but also on the values of free parameters $\alpha $
and $\beta $ (or equivalently $\alpha _{3}$ and $\alpha _{4}$), this
delicate issue is beyond the scope of the current work. Our aim is less
ambitious: \emph{taking the theory with the aforementioned reference metric,
which has been considered numerous times in the literature, what can we say
about the existence of black hole remnant?}

Indeed, considering the ansatz (\ref{metric}), the reference metric (\ref%
{reference}), and the field equation (\ref{Field equation}), we can obtain
the following exact solution \cite{Ghosh}
\begin{equation}
g\left( r\right) =1-\frac{m_0}{r}+\frac{\Lambda r^{2}}{3}+\gamma
r+\varepsilon ,  \label{BH}
\end{equation}%
where $m_0$ is an integration constant related to the mass of the black
hole, while $\Lambda $, $\gamma $ and $\varepsilon $ are, respectively \cite%
{Ghosh},
\begin{eqnarray}
\Lambda &=&3m^{2}\left( 1+\alpha +\beta \right) ,  \notag \\
\gamma &=&-cm^{2}\left( 1+2\alpha +3\beta \right) ,  \notag \\
\varepsilon &=&c^{2}m^{2}\left( \alpha +3\beta \right) .  \label{CondI}
\end{eqnarray}

It is notable that by considering the term $3m^{2}(1+\alpha +\beta )$\
equals to $\Lambda $, one can see that there is a similarity between the
obtained black hole solutions in Eq. (\ref{BH}), and the AdS Schwarzschild
black holes. So we can consider the term $3m^{2}(1+\alpha +\beta )$\ as the
effective cosmological constant. Here we see that the cosmological constant
is \textquotedblleft emergent\textquotedblright\ -- it comes from the
nonzero graviton mass $m$. It follows directly from Eq. (\ref{BH}) that the
Schwarzschild solution is recovered for vanishing massive terms ($m^{2}=0$).
Asymptotically locally anti-de Sitter (AdS)-like and de Sitter (dS)-like
solutions are possible (depending on the sign of $\left( 1+\alpha +\beta
\right) $); for nonzero $\gamma $ and $\varepsilon $ the asymptotic
geometries are not strictly AdS or dS. The constant term $\varepsilon $
corresponds to global monopole \cite{Ghosh}.

Now, we briefly discuss the geometrical structure of this solution. For this
purpose, we first look for the obvious singularity (if any) by studying two
scalar curvatures: Ricci and Kretschmann scalars. Considering the metric (%
\ref{metric}), with the solution (\ref{BH}), the Ricci scalar is given by
\begin{equation}
R=-4\Lambda -\frac{6\gamma }{r}-\frac{2\left( 1+\gamma \right) }{r^{2}}.
\end{equation}

Evidently, we encounter a divergence of the Ricci scalar at the origin ($%
\lim_{r\rightarrow 0}R=\infty $). Also, the Kretschmann scalar ($R_{\alpha
\beta \gamma \delta }R^{\alpha \beta \gamma \delta }$) is given by a rather
lengthy expression:
\begin{eqnarray}
R_{\alpha \beta \gamma \delta }R^{\alpha \beta \gamma \delta } &=&\frac{%
8\Lambda ^{2}}{3}+\frac{8\Lambda \gamma }{r}+\frac{8\left( \gamma ^{2}+\frac{%
\Lambda }{3}\left( 1+\varepsilon \right) \right) }{r^{2}}  \notag \\
&&  \notag \\
&&+\frac{8\gamma \left( 1+\varepsilon \right) }{r^{3}}+\frac{4\left(
1+\varepsilon \right) ^{2}}{r^{4}}  \notag \\
&&  \notag \\
&&-\frac{8m_{0}\left( 1+\varepsilon \right) }{r^{5}}+\frac{12m_{0}^{2}}{r^{6}%
}.
\end{eqnarray}%
\bigskip

One can show that this scalar has the following behavior
\begin{eqnarray}
\lim_{r\longrightarrow 0}R_{\alpha \beta \gamma \delta }R^{\alpha \beta
\gamma \delta } &=&\infty , \\
&&  \notag \\
\lim_{r\longrightarrow \infty }R_{\alpha \beta \gamma \delta }R^{\alpha
\beta \gamma \delta } &=&\frac{8\Lambda ^{2}}{3},
\end{eqnarray}%
which can confirm that there is a curvature singularity at $r=0$, and also
the asymptotic behavior of this solution is AdS\footnote{%
For de-Sitter case, since $r$ is bounded, we cannot take $r\rightarrow
\infty $ limit. {{Note that with the metric Eq. (\ref{BH}), $\Lambda
> 0$ corresponds to anti-de Sitter instead of de-Sitter. This can be a
source of confusion, however we follow the form of Eq. (\ref{BH}) which is
widely used in the literature.}}}, since the Kretschmann scalar is $\frac{%
8\Lambda ^{2}}{3}$ at $r\longrightarrow \infty $.

For clarity, we plot the metric function (\ref{BH}) versus $r$ in Fig. (\ref%
{Fig11}) for specific choice of the parameter values. As one can see, there
is a zero to Eq.(\ref{BH}) which corresponds to the event horizon.

%\textbf{It
%is known that the linear theory of massive gravity becomes strongly coupled
%at low energy for small values of the graviton mass \cite{Vainshtein1972}.
%Therefore, one has to looking for an appropriate theory of massive gravity
%with nonlinear interactions. Regarding the nonlinear dRGT massive gravity,
%one may concern the cutoff scales of theory. The cutoff scale of a theory
%indicates the scale at which the theory breaks down. As an example, the
%cutoff scale of General Relativity (with massless graviton) is the Planck
%scale. Although the cutoff scale of the massive theory could potentially be
%below the Planck scale, it is shown that \cite{dRham2014} such cutoff is
%above the scale the strong-coupling scale of the theory.}
%%%%%%%%%%%%%%%%%%%%%%%%%%%%%%%%%%%%%%%%%%%%%%%%%%%%%%%%%%%%%%%
\begin{figure}[tbp]
$%
\begin{array}{c}
\epsfxsize=7cm \epsffile{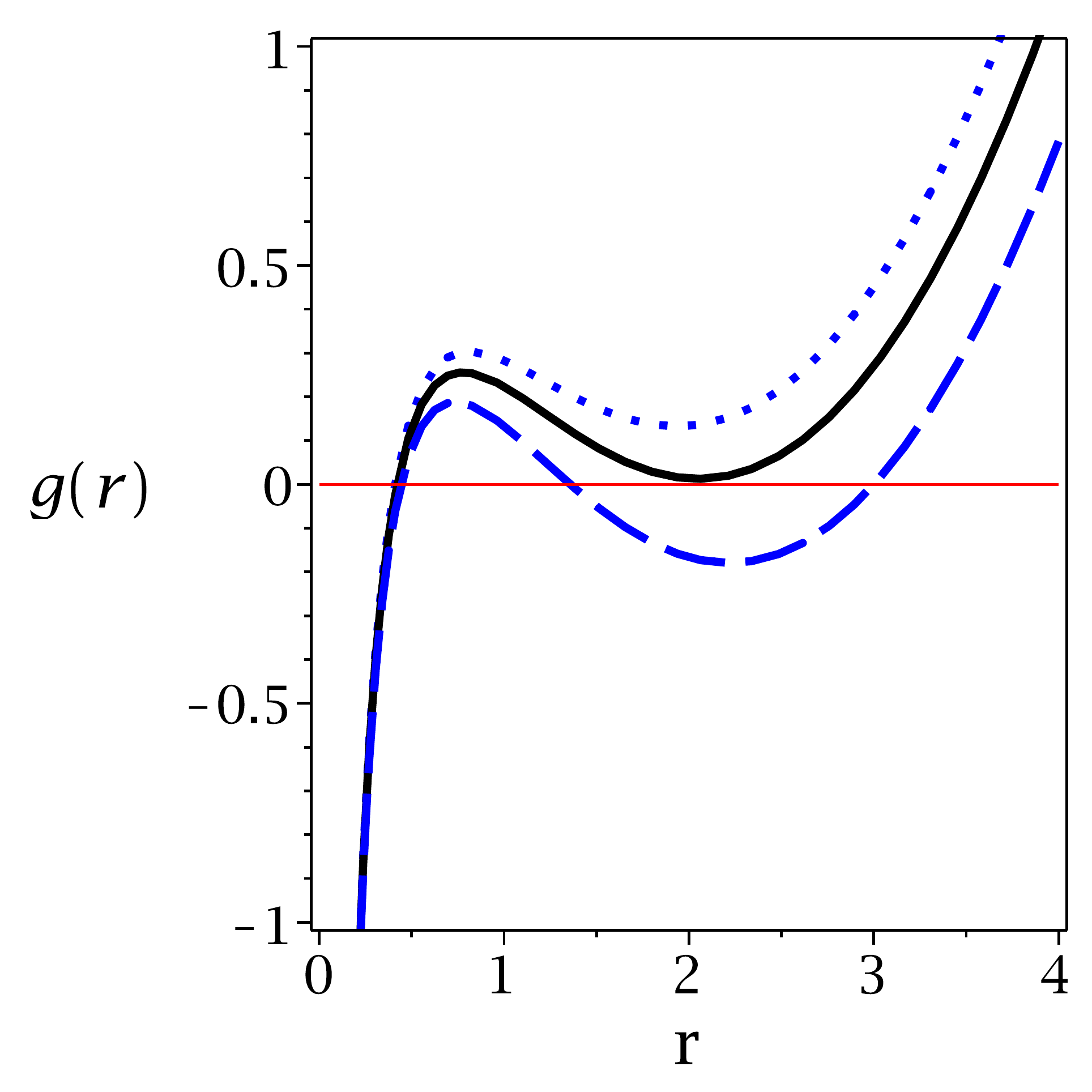}%
\end{array}
$%
\caption{{The plot of $g(r)$ versus $r$, for definiteness here we set
$\Lambda =1$, $\protect\varepsilon =1$, $m_{0}=0.6$, $\protect\gamma=-1.45$
(dotted line), $\protect\gamma=-1.51$ (continuous line) and $\protect\gamma%
=-1.60$ (dashed line).}}
\label{Fig11}
\end{figure}
%%%%%%%%%%%%%%%%%%%%%%%%%%%%%%%%%%%%%%%%%%%%%%%%%%%%%%%%%%%%%%%

In this work we will show that for black hole remnants to exist, it must be
the case that $\Lambda>0$, which, as we have emphasized, from the metric in
Eq. (\ref{BH}), actually corresponds to \emph{anti-de Sitter} case.

%We note that by consider $\alpha =-(1+\beta )$, Eq. (\ref{BH}) reduces to
%\begin{equation}
%g\left( r\right) =1-\frac{2m_{0}}{r}+\gamma r+\varepsilon ,  \label{BHVegh}
%\end{equation}%
%where $\alpha _{3}=-\frac{2+\beta }{3}$ and $\alpha _{4}=\frac{1}{6}\left(
%1+2\beta \right) $; or equivalently, $\gamma $ and $\varepsilon $ are
%given as
%\begin{eqnarray}
%\gamma &=&cm^{2}\left( 1-\beta \right) ,  \nonumber \\
%\varepsilon &=&-c^{2}m^{2}\left( 1-2\beta \right) .
%\end{eqnarray}
%Thus, we observe that by setting $\alpha =-(1+\beta )$, the obtained
%solution in Eq. (\ref{BH}) is the same with the black hole solution obtained by Vegh (Eq. (\ref{BH3}), in \textit{Appendix I}).
The physical mass of the black hole is $M={m_{0}}/{2}$ \cite{Cai}, which can
be obtained from the Hamiltonian method. For $m=0$, $M$ reduces to the
standard ADM mass of an asymptotically flat Schwarzschild black hole. In
order to find the event horizon of black hole, we\textbf{\ }should solve $%
g(r)=0$. This gives rise to a cubic equation with at most three real roots.
For more details see Fig.(\ref{Fig11})\textbf{\ } In fact, the largest real
positive root of the function $g(r)$ is the event horizon of the black hole,
given by
\begin{equation}
r_{+}=\frac{\mathcal{A}^{2/3}-2\gamma \mathcal{A}^{1/3}-4\left(
1+\varepsilon \right) \Lambda +4\gamma ^{2}}{2\Lambda \mathcal{A}^{1/3}},
\label{r+}
\end{equation}%
where
\begin{flalign}
\mathcal{A}=&8\Lambda \left( 1+\varepsilon \right) \left\{\frac{%
3M\left( 3M\Lambda ^{2}-2\gamma ^{3}\right) }{\left( 1+\varepsilon \right)
^{2}}+ \right. \notag \\&\left.\frac{\Lambda \left[ \left( 1+\varepsilon \right) ^{2}+9\gamma M\right]
}{1+\varepsilon }-\frac{3\gamma ^{2}}{4}\right\}^{\frac{1}{2}}
+24M\Lambda^{2}\notag \\&+12\gamma \left( 1+\varepsilon \right) \Lambda -8\gamma
^{3}.
\end{flalign}The above root depends on the values of the various parameters.
In fact, as the black hole evaporates and the mass decreases, it is possible
that the number of real roots changes. Regardless, one identifies the
largest root as the event horizon.

The Hawking temperature can be obtained from Eqs. (\ref{metric}), (\ref{BH})
and (\ref{r+}) using the standard method:
\begin{equation}
T=\frac{1}{4\pi }\left. g{^{\prime }}\left( r\right) \right\vert
_{r=r_{+},m_{0}=2M}=\frac{1+\Lambda r_{+}^{2}+2\gamma r_{+}+\varepsilon }{%
4\pi r_{+}}.  \label{TdRGT}
\end{equation}

The Bekenstein-Hawking entropy of the black hole can be calculated through
the first law of black hole thermodynamics $\text{d}S={\text{d}M}/{T}$,
which yields the standard area law upon integration: $S=\pi r_{+}^{2}.
%\label{SdRGT}
$ The heat capacity is
\begin{equation}
C =T\left( \frac{\partial S}{\partial T}\right) =\frac{2\pi \left( 1+\Lambda
r_{+}^{2}+2\gamma r_{+}+\varepsilon \right) r_{+}^{2}}{\Lambda
r_{+}^{2}-(1+\varepsilon )}.  \label{C}
\end{equation}

As the black hole evaporates, it gradually loses its mass. In general
relativity, the temperature of the black hole diverges as $M\rightarrow 0$.
This is true even for black holes in anti-de Sitter spacetime (large enough
black holes however, do not evaporate if we impose the usual reflective
boundary condition at conformal infinity. See refs. \cite{Ong2016,Yao2019},
for more details.) Note that, in this work, we consider a model in which the
cosmological constant arises from graviton mass $m$, therefore once the
parameters $\alpha ,\beta $ are fixed to have nonzero $\Lambda ,\gamma $ and
$\varepsilon $, then general relativity is equivalent to taking $m=0$, i.e.
without a cosmological constant. That is to say, the black hole reduces to
asymptotically flat Schwarzschild black hole, not Schwarzschild-anti de
Sitter black hole.

In various quantum gravity inspired phenomenological models, such as the
generalized uncertainty principle, black holes do not evaporate completely
but instead become a remnant \cite{0106080}.\textbf{\ }Here, we want to show
that the black holes in massive gravity do not evaporate completely and we
end up with a remnant. For this purpose we solve for the real positive root
of the temperature expression set to zero, $T=0$, which is given by
following form:\textbf{\ }%
\begin{equation}
R_{r}=\frac{\sqrt{\gamma ^{2}-(1+\varepsilon )\Lambda }-\gamma }{\Lambda }.
\label{Rr}
\end{equation}%
The square root actually comes with a $\pm $ sign. However, for
definiteness, let us consider the sign in front of the square root to be
positive. A black hole remnant also exists if we choose the negative sign.

By substituting $R_{r}$\ in the physical mass $M_{r}\left( r=R_{r}\right) $,
one finds that indeed there exists a remnant with mass below which there is
no black hole solution:
\begin{eqnarray}
M_{r} &=&\frac{2\Lambda (1+\varepsilon )+\gamma \left( \sqrt{\gamma
^{2}-(1+\varepsilon )\Lambda }-\gamma \right) }{6\Lambda ^{2}}  \notag \\
&&  \notag \\
&&\times \left( \sqrt{\gamma ^{2}-(1+\varepsilon )\Lambda }-\gamma \right) .
\label{Mr}
\end{eqnarray}

%%%%%%%%%%%%%%%%%%%%%%%%%%%%%%%%%%%%%%%%%%%%%%%%%%%%%%%%%%%%%%%
\begin{figure}[tbp]
$%
\begin{array}{cc}
\epsfxsize=4.35cm \epsffile{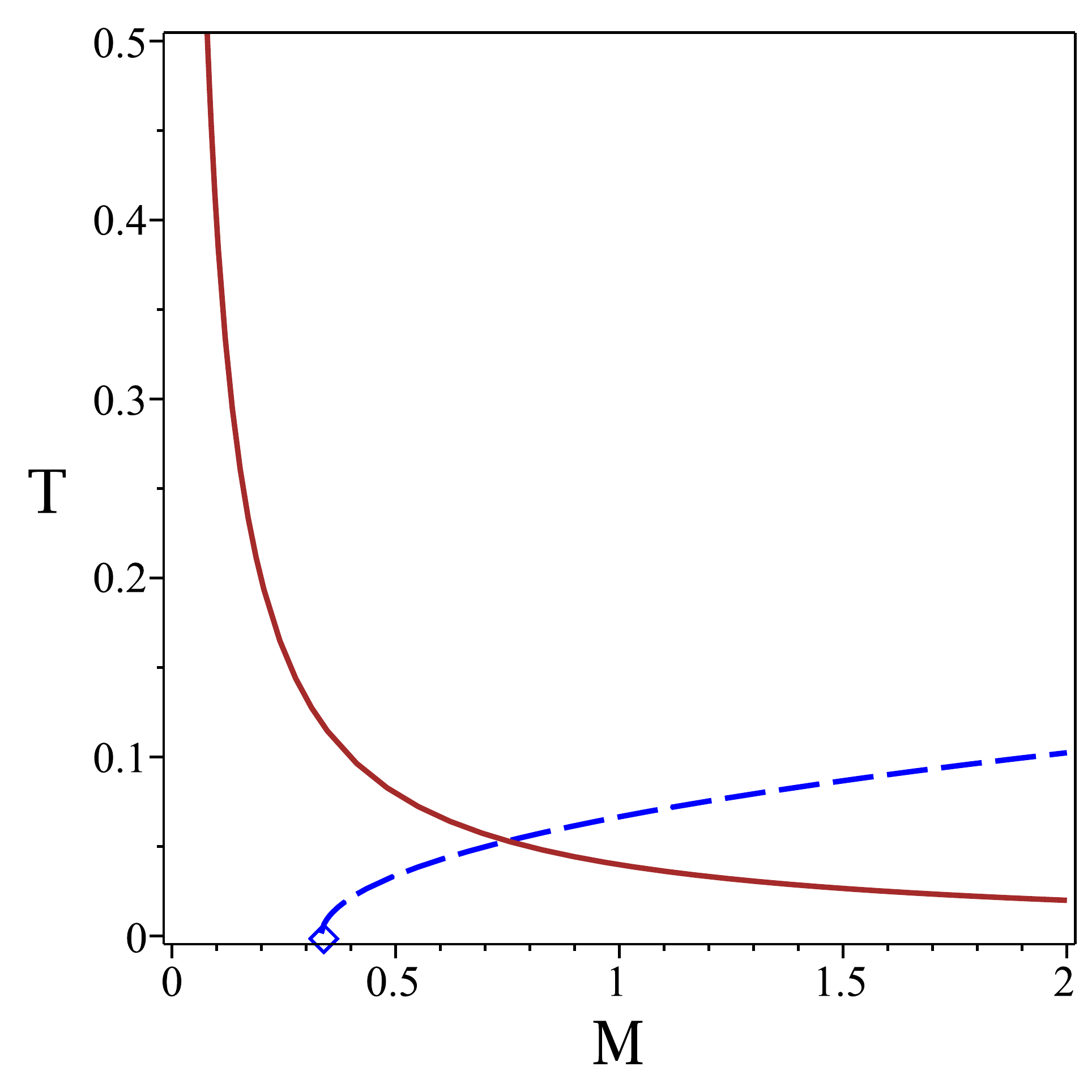} & \epsfxsize=4.35cm \epsffile{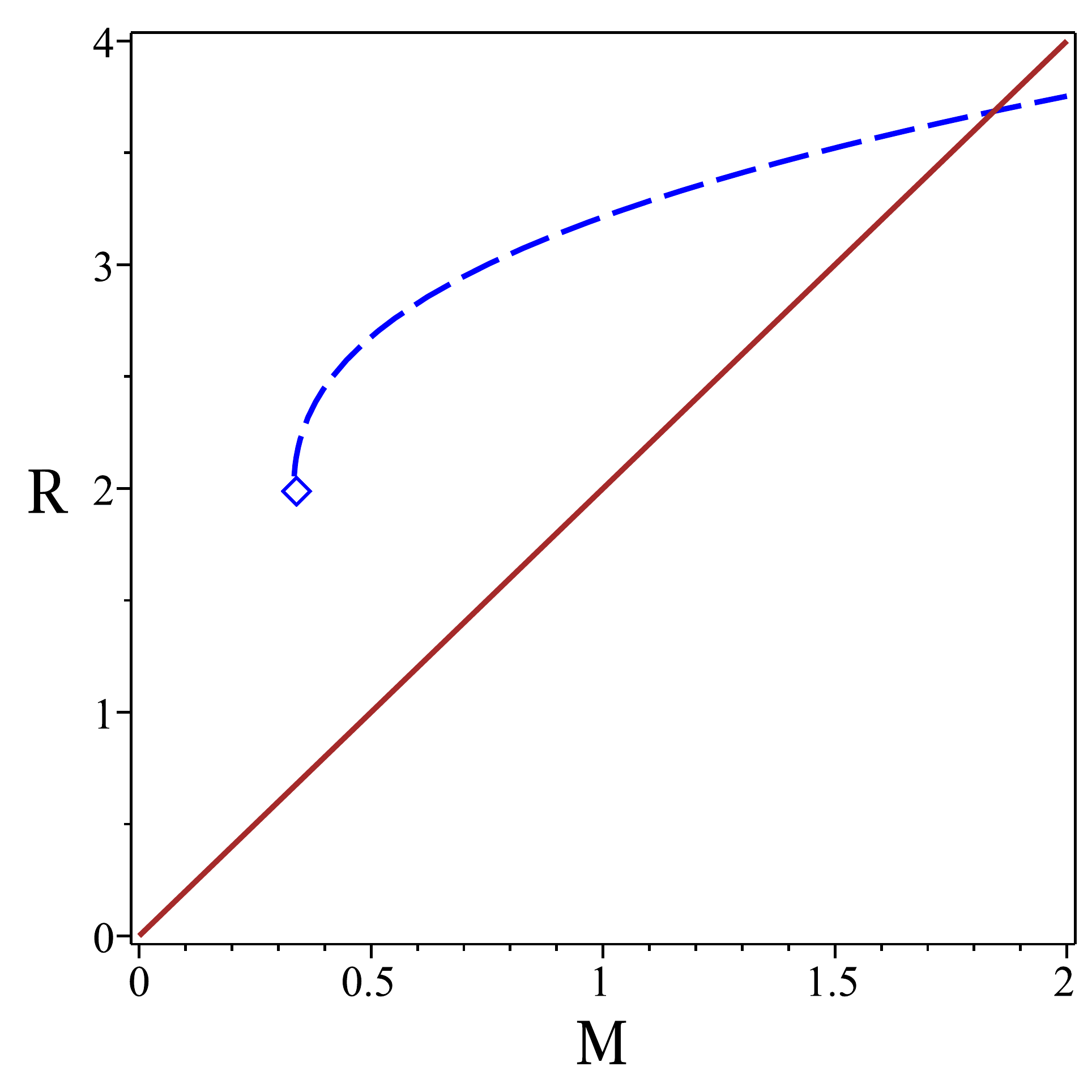} \\
\epsfxsize=4.35cm \epsffile{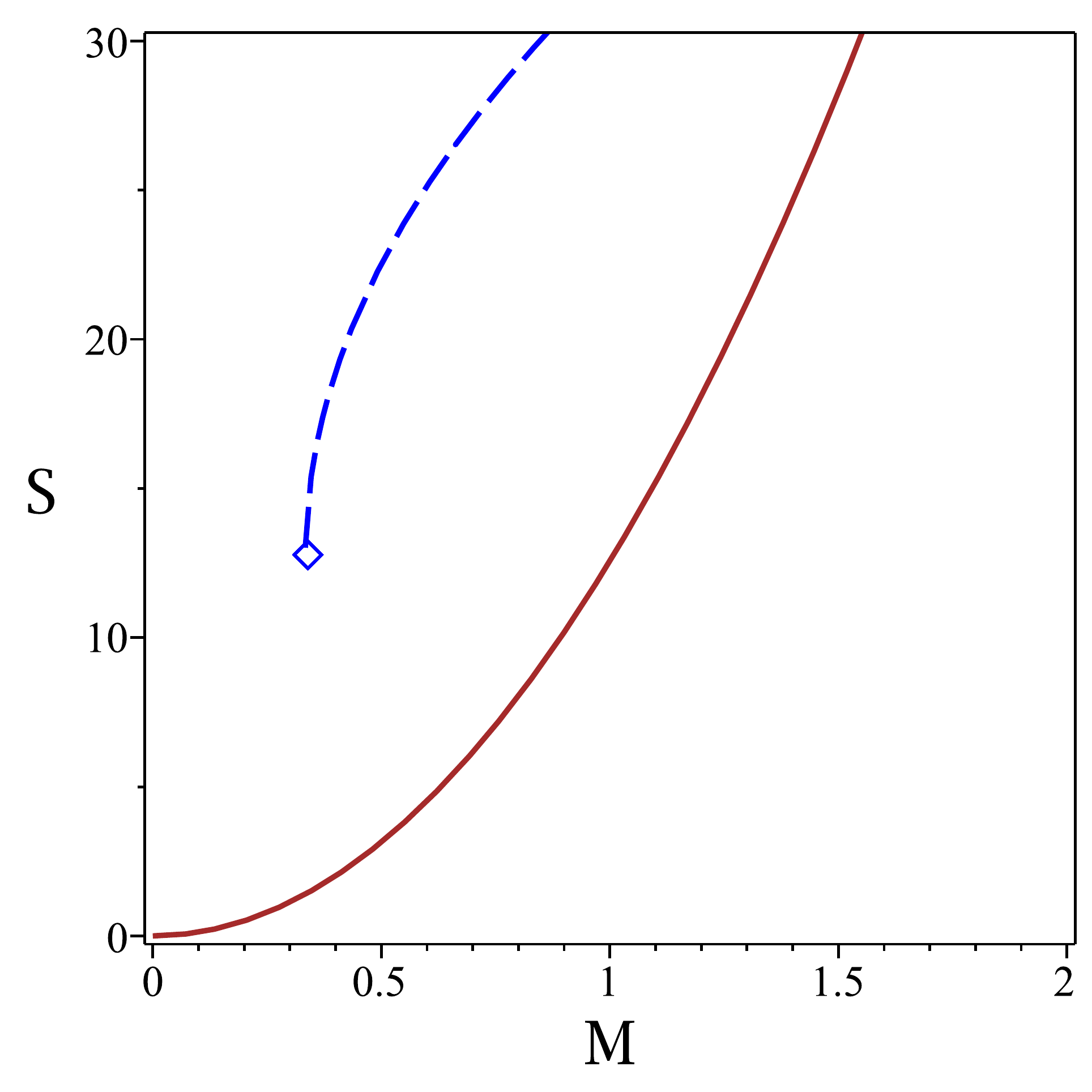} & \epsfxsize=4.35cm \epsffile{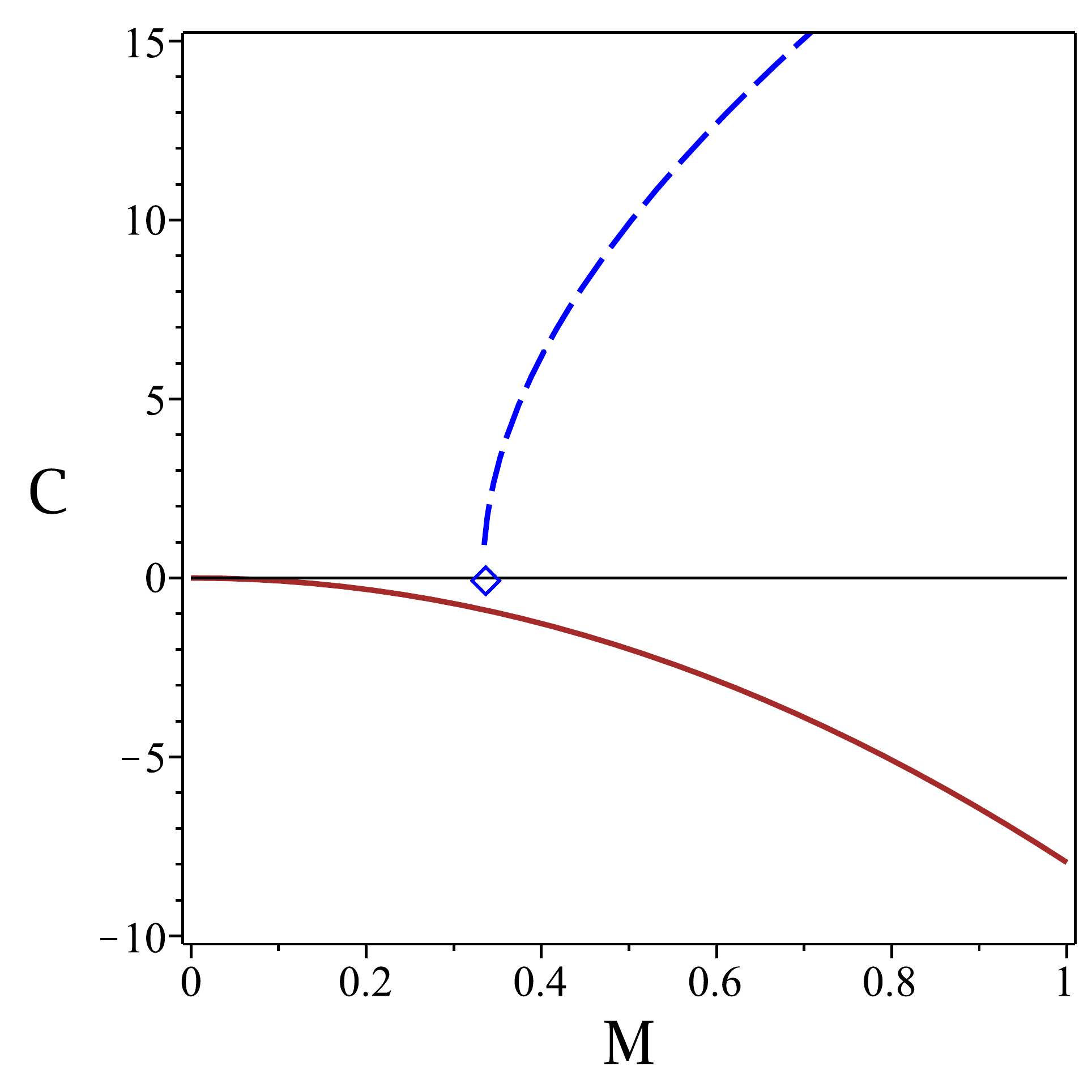}%
\end{array}
$%
\caption{For this illustrative example we take $\Lambda =1$, $\protect%
\varepsilon =1$, $\protect\gamma =-1.5$. General relativity ($m=0$) yields
asymptotically flat Schwarzschild black hole. Dashed and continuous lines
are related to black holes in massive gravity and general relativity,
respectively. \textbf{Up left panel:} Temperature ($T$) versus mass ($M$).
\textbf{Up right panel:} Radius ($R$) versus mass ($M$). \textbf{Down left
panel:} Entropy ($S$) versus mass ($M$). \textbf{Down right panel:} Heat
capacity ($C$) versus mass ($M$).}
\label{Fig1}
\end{figure}
%%%%%%%%%%%%%%%%%%%%%%%%%%%%%%%%%%%%%%%%%%%%%%%%%%%%%%%%%%%%%%%

We remark that in the absence of the cosmological constant the remnant mass
of black hole reduces to
\begin{equation}
M_{r}(\Lambda =0)=-\frac{(1+\varepsilon )^{2}}{8\gamma }.  \label{Mrr}
\end{equation}%
In order to have a positive remnant mass of the black hole in the absence of
the cosmological constant (Eq. (\ref{Mrr})), we always impose $\gamma <0$.
%or equivalently, $\beta >1$.

The remnant of entropy is given by
\begin{equation}
S_{r}=\frac{\pi }{\Lambda ^{2}}\left( \sqrt{\gamma ^{2}-(1+\varepsilon
)\Lambda }-\gamma \right) ^{2}.  \label{Srr}
\end{equation}

We present our results in Fig. (\ref{Fig1}), in which the temperature,
radius, entropy and heat capacity of a massive gravity black hole, with
parameters chosen to be $\Lambda =1=\varepsilon ,\gamma =-1.5$, are depicted
and compared against its GR counter-part. We note in particular that the
heat capacity and temperature are both zero when the remnant mass is
reached. Recall that for $\Lambda >0$, our black hole
remnant is asymptotically anti-de Sitter.

{One might wonder if }$\Lambda <0$ {\ solution also exists,
which might be more relevant to cosmology? Unfortunately this seems not to
be the case.} {To see this, let us consider Eq. (\ref{Rr}) and Eq. (\ref{Mr}). We will show that for }$\Lambda
<0${\ (which would correspond to the de Sitter case), it is
impossible to obtain remnants for which the mass and the radius are positive
simultaneously. With }$\Lambda <0${, in Eqs. (%
\ref{Rr}) and (\ref{Mr}) can be written as }%
\begin{eqnarray}
R_{r} &=&\frac{\sqrt{\gamma ^{2}+\sigma }-\gamma }{\Lambda },  \label{R} \\
&&  \notag \\
M_{r} &=&\frac{\left[ -2\sigma +\gamma \left( \sqrt{\gamma ^{2}+\sigma }%
-\gamma \right) \right] \left( \sqrt{\gamma ^{2}+\sigma }-\gamma \right) }{%
6\Lambda ^{2}},  \label{M}
\end{eqnarray}%
{in the above equation }$\sigma =-(1+\varepsilon )\Lambda $\textbf{.}

{In order to have a real positive remnant radius (}$R_{r}>0${)
for }$\Lambda <0${, the following constraints must be satisfied: }%
\begin{eqnarray}
\sqrt{\gamma ^{2}+\sigma }-\gamma &<&0,  \label{I} \\
&&  \notag \\
\gamma ^{2}+\sigma &>&0,  \label{II}
\end{eqnarray}%
\{in which due to Eq. (\ref{I}), we find the requirement that $\sigma
<0${\ (or }$\varepsilon <-1$){. }

{Considering the condition (\ref{II}), we obtain two possible ranges
for }$\gamma ${: either }$\gamma \geqslant \sqrt{-\sigma }${\
or }$\gamma \leqslant -\sqrt{-\sigma }${. However, note that Eq. (\ref%
{I}) implies that }$\gamma >0${, so we must have }$\gamma \geqslant
\sqrt{-\sigma }${. }

\{Now, we focus on the remnant mass of black holes given by Eq. (\ref%
{M}). In order to have positive value of remnant mass, the numerator of Eq. (%
\ref{M}) has to be positive, which contains two factors:

\begin{itemize}
\item[(I)] $-2\sigma +\gamma \left( \sqrt{\gamma ^{2}+\sigma }-\gamma
\right) ${, \newline
}

\item[(II)] $\sqrt{\gamma ^{2}+\sigma }-\gamma ${. \newline
}
\end{itemize}

{They must either be both positive or both negative. However, factor
(II) is nothing but }$-\Lambda R_{r}${, which we want to be positive
(for }$\Lambda <0${). According to Eq. (\ref{I}), the factor (II) is
negative. It thus follows that we need factor (I) to also be negative. Since
previously }$\sigma <0${, the term }$-2\sigma ${\ in factor
(I) is positive. The question is whether the factor (I) can be negative enough. The
answer is no. In order to have the negative value for the factor (I), we
need }$\gamma \left( \sqrt{\gamma ^{2}+\sigma }-\gamma \right)
<2\sigma ${. Since }$\gamma \geqslant \sqrt{-\sigma }${, we
define }$\gamma :=s\sqrt{-\sigma }${\ in which }$s\geqslant 1${%
. Replacing }$\gamma =s\sqrt{-\sigma }${\ in }$\gamma \left( \sqrt{%
\gamma ^{2}+\sigma }-\gamma \right) <2\sigma ${, we obtain }%
\begin{equation}
s\sigma \left( s-\sqrt{s^{2}-1}\right) <2\sigma .  \label{S}
\end{equation}
{{
Since $\sigma <0$, this means we need
\begin{equation}
s \left( s-\sqrt{s^{2}-1}\right) > 2,  
\end{equation}
which yields $s < -2\sqrt{3}/3$, contradicting $s \geqslant 1$.
This shows that there is no black hole remnant for }}$\Lambda <0${\ (or black hole remnant
which is asymptotically de Sitter).}

%%%%%%%%%%%%%%%%%%%%%%%%%%%%%%%%%%%%%%%%%%%%%%%%%%%%%%%%%%%%%%%%%%%%%%%%%%%%%%%%%%%%%%%%%

\section{Black Hole Remnant and Information Paradox in Anti-de Sitter-Like
Universe}

There are a few motivations for black hole remnants, one of which is that
remnants prevent black holes from becoming arbitrarily hot during the end
stage of the evaporation. Probably no one expects Hawking temperature to be
truly divergent in the $M \to 0$ limit, but exactly what prevents just such
a divergence is not agreed upon. One possibility is simply that new physics
comes in at sufficiently high energy, thereby stopping black holes from
evaporating further. Just such a possibility was investigated in \cite%
{0106080} by appealing to the generalized uncertainty principle (GUP), which
modifies quantum mechanics taking into account correction due to strong
gravitational field. The remnant solution therein exhibits a rather peculiar
property that its temperature is positive -- how could a black hole be a
remnant (not losing mass) yet continue to have Hawking radiation? One
possible way out is to interpret this temperature as the internal energy of
the remnant (since $E\sim kT$).
%which is rather like an elementary particle
The specific heat of the remnant is zero, and therefore it does not interact
with the thermal environment \cite{0912.2253, 1806.03691}. This means the
remnant is stable, a pre-requisite for it to serve as dark matter candidate.
Indeed, such a black hole remnant derived from GUP has been proposed as
possible dark matter candidate \cite{0205106}.

Another virtue of black hole remnant is that it might be able to ameliorate
the information paradox of black hole. The usual proposal to preserve
quantum information is by having it scrambled and entangled in the Hawking
radiation. Consider a black hole formed by a pure state. By unitarity one
should recover pure state at the end of the black hole evaporation. The
attempt to purify the Hawking radiation has given rise to issues like
firewall \cite{1207.3123}. The remnant picture, first proposed in \cite{ACN}%
, avoided this problem by proposing that Hawking radiation is never purified
-- states behind the horizon and states in the Hawking radiation remains
mixed separately, but taken as a whole it is a pure state. Such a proposal
is not without problems. For example, in order to hide plenty of quantum
states behind the ever shrinking horizon, the Bekenstein-Hawking entropy
does not reflect all the interior degrees of freedom. There is also the
infinite production problem. Both of these problems are discussed in details
in \cite{Chen}. The bottom line is that despite these issues, remnants
should not be dismissed outright, and could well help to resolve the
information paradox, especially if they have huge interiors due to
non-trivial geometries. All these comments apply also to our massive gravity
remnants, with the caveats that our remnants exist in anti-de Sitter
spacetime.

There are two ways in which our remnants can be relevant to the information
paradox. The first possibility is more straightforward: as we have explained
in the introduction, our Universe could actually be asymptotically anti-de
Sitter in the far future, with the current phase of acceleration caused by
other fields \cite{0403104} (This would also avoid the recently raised
\textquotedblleft swampland\textquotedblright\ issue of de Sitter space \cite%
{1806.08362}.). In \cite{0403104}, a quintessence was used. It is now
appreciated that a simple quintessence model is difficult to be realized in
string theory without fine tuning, so appealing to one to avoid the
Swampland is swapping one difficulty with another \cite{1808.09440,1811.05434,
1808.08967, 1812.03184}. However, other more complicated fields could still
do the job \cite{1902.11014v2}.

Of course, there is the subtlety that the theory still needs to be coupled
with these other fields and then strictly speaking the remnant solution
would be different (if it still exists). If our Universe is asymptotically
de Sitter, as most cosmologists believe, then our massive gravity remnants
cannot be straightforwardly applied to understand actual black holes.
Nevertheless, it is hoped that black hole remnants in the anti-de Sitter
bulk may -- eventually -- help us to understand how information is preserved
via holographic correspondence to a field theory on the conformal boundary.

The thermal stability of the massive gravity remnant is demonstrated by the
fact that the heat capacity is zero, much like the remnant obtained from GUP
mentioned above (this is not always the case for all GUP models, see e.g.,
\cite{1806.03691}). This means that we have a thermodynamically inert and
stable remnant. However, unlike the GUP remnant, its temperature is also
zero. This is in fact much more natural -- no mass is loss via Hawking
emission and thus the remnant is stable. While we could argue that the GUP
remnant temperature is really its internal energy, this feels somewhat
contrived in comparison. Since our black hole has no electrical charge, the
remnant is not like an extremal charged black hole, which could continue to
radiate (despite having zero temperature) via non-thermal processes such as
Schwinger pair production \cite{Chen}; the remnant is arguably more stable
and long-lived.
%%%%%%%%%%%%%%%%%%%%%%%%%%%%%%%%%%%%%%%%%%%%%%%%%%%%%%%%%%%%%%%%%%%%%

\section{Discussions}

In this work we investigated whether dRGT massive gravity can admit remnant
scenario, and found that it is indeed possible. To our knowledge, this is
the first black hole remnant found in dRGT massive gravity. The black hole
tends to zero temperature remnant with vanishing specific heat, at which
point it stops evaporating and becomes stable. The remnant only has positive
mass for $\gamma <0$. Massive gravity remnant could help to ameliorate the
information paradox, modulo the usual challenges \cite{Chen}. Here we
discuss several issues and outlook for future works. Note that in \cite%
{1610.01505}, a solution in dyonic massive gravity was discussed in which
there is a \textquotedblleft remnant temperature\textquotedblright , i.e. in
the limit of vanishing radius, the temperature is nonzero -- it is not a
black hole remnant in the sense studied in this work.

In this work, we chose the reference metric
\begin{equation}
f_{ab}=\text{diag}(0,0,c^{2},c^{2}\sin ^{2}\theta ).
\end{equation}

Since different reference metric might give different results, a more
detailed analysis is required to find out how our results may change if
another reference metric is chosen. In particular, although our result shows
that remnants can only exist in asymptotically anti-de Sitter spacetime,
other reference metric may allow remnants to exist in asymptotically de
Sitter spacetime as well. This will require further investigations.

It is worth commenting on the cutoff scale of massive gravity theory. The
cutoff scale of any theory is the scale beyond which the theory breaks down
(in the sense of effective field theory), and one would expect new physics
to come in at higher energy scale. The cutoff scale of general relativity
(with massless graviton) is the Planck scale. As mentioned in \cite{Rham},
the cutoff scale for massive gravity is not the same as the strong-coupling
scale of the theory. Furthermore, the latter need not necessarily mean that
there is new physics, but only that perturbation theory breaks down. In fact
strong gravity tends to raise the strong coupling scale \cite{1709.07503}.
Except from very near the singularities, black hole solutions we discussed
here are valid in massive gravity theory.

Demonstrating that dRGT massive gravity admits black hole remnant solutions
is only the first step in the analysis. One needs to consider the actual
evolution of the black hole under Hawking evaporation. That is to say, one
has to study the mass loss rate $\text{d}M/\text{d}t$. The importance of
doing so is to check if the remnant state is attainable, i.e. if it can be
reached in a finite time, such an analysis would be important to study the
Page time \cite{9305007, 9306083, 1301.4995} of the black hole. (Conversely,
even if there is no remnant, one could have an \textquotedblleft effective
remnant\textquotedblright\ if the evaporation rate is infinite \cite%
{1806.03691}.) {Presumably if the third law of black hole thermodynamics is
valid for such black hole, it would take infinite amount of time to reach
zero temperature state.} In addition to the mass loss rate $\text{d}M/\text{d%
}t$, one should also study the sparsity of the Hawking radiation \cite%
{1806.03691,1506.03975,1512.05809}, which affects the lifetime of the black
hole. This is beyond the scope of the current paper, and is left for future
works.

As mentioned in Sec.(\ref{s1}), dRGT massive gravity suffers from a variety
of problems, most notably the causality issue which plagues the theory with
superluminal propagation and arbitrarily small closed causal curves, thus
rendering the theory {rather unpredictive}. In addition, a
\textquotedblleft god-given\textquotedblright\ reference metric is somewhat
unsatisfactory. These has led to the considerations of bimetric
(Hassan-Rosen) theory \cite{1109.3515}, in which the reference metric $%
f_{ab} $ is dynamical. Such a theory has some advantages over the original
massive gravity \cite{1503.07521}, and its causal structures and constraints
are gradually being understood \cite{1706.07806, 1802.07267}, though more
research is clearly needed.

Finally, let us speculate on the possibility that massive gravity remnants
may be dark matter candidate. Black hole remnants as dark matter is of
course not a new idea, see, e.g. \cite{MacGibbon, Barrow, Carr} for some
early examples. If our Universe is fundamentally anti-de Sitter, which the
current phase of accelerated expansion caused by another field, say a
quintessence, then it is possible that massive gravity remnants may play a
role as dark matter. In addition, the idea that massive gravitons might be
dark matter themselves had been proposed quite a few years back \cite%
{0411158}. Massive gravitons remain possible as dark matter candidate in the
context of bimetric gravity \cite{1604.06704, 1604.08564, 1607.03497,
1708.04253}. If remnants exist in that theory they could serve as an
additional dark matter candidate.

%%%%%%%%%%%%%%%%%%%%%%%%%%%%%%%%%%%%%%%%%%%%%%%%%%%%%%%%%%%%%

\begin{acknowledgments}
We thank an anonymous referee for useful comments. BEP and SHH wish to thank
the Shiraz University Research Council. The work of BEP has been supported
financially by the Research Institute for Astronomy and Astrophysics of
Maragha (RIAAM) under research project No. 1/5750-57. YCO thanks the
National Natural Science Foundation of China (grant No.11705162) and the
Natural Science Foundation of Jiangsu Province (No.BK20170479) for funding
support. YCO also thanks Nordita, where this work was carried out, for
hospitality during his summer visit for the Lambda Program. YCO also thank
members of Center for Gravitation and Cosmology (CGC) of Yangzhou University
(\hypersetup{urlcolor=purple}{%
\href{http://www.cgc-yzu.cn}{http://www.cgc-yzu.cn}}) for discussions.
\end{acknowledgments}


\begin{thebibliography}{99}
\bibitem{Giddings} Steven B. Giddings, {\hypersetup{urlcolor=vividviolet}
\href{https://journals.aps.org/prd/abstract/10.1103/PhysRevD.46.1347}{Phys.
Rev. D \textbf{46} (1992) 1347}}, \href{https://arxiv.org/abs/hep-th/9203059}%
{[arXiv:hep-th/9203059]}.

\bibitem{Giddings2} Steven B. Giddings, ``Black Holes and Massive
Remnants'', {\hypersetup{urlcolor=vividviolet}\href{https://journals.aps.org/prd/abstract/10.1103/PhysRevD.46.1347}%
{Phys. Rev. D \textbf{49} (1994) 4078}}, \href{https://arxiv.org/abs/hep-th/9203059}%
{[arXiv:hep-th/9203059]}.

\bibitem{Chen} Pisin Chen, Yen Chin Ong, Dong-han Yeom, ``Black Hole
Remnants and the Information Loss Paradox'', {%
\hypersetup{urlcolor=vividviolet}\href{https://www.sciencedirect.com/science/article/pii/S0370157315004391?via\%3Dihub}%
{Phys. Rep. \textbf{603} (2015) 1}}, \href{https://arxiv.org/abs/1412.8366}{%
[arXiv:1412.8366 [gr-qc]]}.

\bibitem{DAmico} Guido D'Amico, Claudia de Rham, Sergei Dubovsky, Gregory
Gabadadze, David Pirtskhalava, Andrew J. Tolley, \textquotedblleft Massive
Cosmologies\textquotedblright , {\hypersetup{urlcolor=vividviolet}\href{https://journals.aps.org/prd/abstract/10.1103/PhysRevD.84.124046}%
{Phys. Rev. D \textbf{84} (2011) 124046}}, \href{https://arxiv.org/abs/1108.5231}%
{[arXiv:1108.5231 [hep-th]]}.

%\bibitem{Arkani2002} Nima Arkani-Hamed, Savas Dimopoulos, Gia Dvali, Gregory
%Gabadadze, \textquotedblleft Non-Local Modification of Gravity and the
%Cosmological Constant Problem\textquotedblright , {%
%\hypersetup{urlcolor=vividviolet}} \href{https://arxiv.org/abs/hep-th/0209227}%
%{[arXiv:hep-th/0209227]}.

%\bibitem{Dvali2007} Gia Dvali, Stefan Hofmann, Justin Khoury,
%\textquotedblleft Degravitation of the cosmological constant and graviton
%width\textquotedblright , {\hypersetup{urlcolor=vividviolet}\href{https://journals.aps.org/prd/abstract/10.1103/PhysRevD.76.084006}%
%{Phys. Rev. D \textbf{76} (2007) 084006}}, \href{https://arxiv.org/abs/hep-th/0703027}%
%{[arXiv:hep-th/0703027]}.

%\bibitem{Rham2008} Claudia de Rham, Gia Dvali, Stefan Hofmann, Justin
%Khoury, Oriol Pujolas, Michele Redi, Andrew J. Tolley, \textquotedblleft
%Cascading Gravity: Extending the Dvali-Gabadadze-Porrati Model to Higher
%Dimension\textquotedblright , {\hypersetup{urlcolor=vividviolet}\href{https://journals.aps.org/prl/abstract/10.1103/PhysRevLett.100.251603}%
%{Phys. Rev. Lett \textbf{100} (2008) 251603}}.

%\bibitem{Berezhiani} Zurab Berezhiani, Fabrizio Nesti, Luigi Pilo, Nicola
%Rossi, \textquotedblleft Gravity modification with Yukawa-type potential:
%dark matter and mirror gravity\textquotedblright , {%
%\hypersetup{urlcolor=vividviolet}\href{https://iopscience.iop.org/article/10.1088/1126-6708/2009/07/083}%
%{JHEP \textbf{07} (2009) 083}}, \href{https://arxiv.org/abs/0902.0144v2}{%
%[arXiv:0902.0144 [hep-th]]}.

\bibitem{BoundIII} Benjamin P. Abbott et al. (LIGO Scientific Collaboration
and Virgo Collaboration), ``Observation of Gravitational Waves from a Binary
Black Hole Merger'', {\hypersetup{urlcolor=vividviolet}\href{https://journals.aps.org/prl/abstract/10.1103/PhysRevLett.116.061102}%
{Phys. Rev. Lett. \textbf{116} (2016) 061102}}, \href{https://arxiv.org/abs/1602.03837}%
{[arXiv:1602.03837 [gr-qc]]}.

\bibitem{BoundIV} Benjamin P. Abbott et al. (LIGO Scientific and Virgo
Collaborations), ``Tests of General Relativity with GW150914'', {%
\hypersetup{urlcolor=vividviolet}\href{https://journals.aps.org/prl/abstract/10.1103/PhysRevLett.116.221101}%
{Phys. Rev. Lett. \textbf{116} (2016) 221101}}, \href{https://arxiv.org/abs/1602.03841}%
{[arXiv:1602.03841 [gr-qc]]}.

%\bibitem{Bound} Lee Samuel Finn, Patrick J. Sutton, ``Bounding the Mass of
%the Graviton Using Binary Pulsar Observations'', {%
%\hypersetup{urlcolor=vividviolet}\href{https://journals.aps.org/prd/abstract/10.1103/PhysRevD.65.044022}%
%{Phys. Rev. D \textbf{65} (2002) 044022}}, \href{https://arxiv.org/abs/gr-qc/0109049}%
%{[arXiv:gr-qc/0109049]}.

%\bibitem{BoundI} Andrei Gruzinov, ``On the Graviton Mass'', {%
%\hypersetup{urlcolor=vividviolet}\href{https://www.sciencedirect.com/science/article/pii/S1384107605000023?via\%3Dihub}%
%{New Astron. \textbf{10} (2005) 311}}, \href{https://arxiv.org/abs/astro-ph/0112246}%
%{[arXiv:astro-ph/0112246]}.

%\bibitem{BoundII} Alfred Scharff Goldhaber, Michael Martin Nieto, ``Photon
%and Graviton Mass Limits'', {\hypersetup{urlcolor=vividviolet}\href{https://journals.aps.org/rmp/abstract/10.1103/RevModPhys.82.939}%
%{Rev. Mod. Phys. \textbf{82} (2010) 939}}, \href{https://arxiv.org/abs/0809.1003}%
%{[arXiv:0809.1003 [hep-ph]]}.

%\bibitem{BoundV} Jose Beltr\'an Jim\'enez, Federico Piazza, Hermano Velten,
%``Evading the Vainshtein Mechanism with Anomalous Gravitational Wave Speed:
%Constraints on Modified Gravity from Binary Pulsars'', {%
%\hypersetup{urlcolor=vividviolet}\href{https://journals.aps.org/prl/abstract/10.1103/PhysRevLett.116.061101}%
%{Phys. Rev. Lett. \textbf{116} (2016) 061101}}, \href{https://arxiv.org/abs/1507.05047}%
%{[arXiv:1507.05047 [gr-qc]]}.

%\bibitem{BoundVI} Claudia de Rham, J. Tate Deskins, Andrew J. Tolley,
%Shuang-Yong Zhou, ``Graviton Mass Bounds'', {%
%\hypersetup{urlcolor=vividviolet}\href{https://journals.aps.org/rmp/abstract/10.1103/RevModPhys.89.025004}%
%{Rev. Mod. Phys. \textbf{89} (2017) 025004}}, \href{https://arxiv.org/abs/1606.08462}%
%{[arXiv:1606.08462 [astro-ph.CO]]}.



\bibitem{Boulware} David G. Boulware, Stanley Deser, ``Can Gravitation Have
a Finite Range?'', {\hypersetup{urlcolor=vividviolet}\href{https://journals.aps.org/prd/abstract/10.1103/PhysRevD.6.3368}%
{Phys. Rev. D \textbf{6} (1972) 3368}}.

\bibitem{FP} Markus Fierz, Wolfgang Pauli, ``On Relativistic Wave
Equations for Particles of Arbitrary Spin in an Electromagnetic Field'', {%
\hypersetup{urlcolor=vividviolet}\href{http://rspa.royalsocietypublishing.org/content/173/953/211.article-info}%
{Proc. R. Soc. A \textbf{173} (1939) 211}}.

\bibitem{dRGT0} Claudia de Rham, Gregory Gabadadze, ``Generalization of the
Fierz-Pauli Action'', {\hypersetup{urlcolor=vividviolet}\href{https://journals.aps.org/prd/abstract/10.1103/PhysRevD.82.044020}%
{Phys. Rev. D \textbf{82} (2010) 044020}}, \href{https://arxiv.org/abs/1007.0443}%
{[arXiv:1007.0443 [hep-th]]}.

\bibitem{dRGT} Claudia de Rham, Gregory Gabadadze, Andrew J. Tolley,
\textquotedblleft Resummation of Massive Gravity\textquotedblright , {%
\hypersetup{urlcolor=vividviolet}\href{https://journals.aps.org/prl/abstract/10.1103/PhysRevLett.106.231101}%
{Phys. Rev. Lett. \textbf{106} (2011) 231101}}, \href{https://arxiv.org/abs/1011.1232}%
{[arXiv:1011.1232 [hep-th]]}.

\bibitem{Rham} Claudia de Rham, \textquotedblleft Massive
Gravity\textquotedblright , {\hypersetup{urlcolor=vividviolet}\href{https://link.springer.com/article/10.12942/lrr-2014-7}%
{Living Rev. Relativ. \textbf{17} (2014) 7}}, \href{https://arxiv.org/abs/1401.4173}%
{[arXiv:1401.4173 [hep-th]]}.

\bibitem{Hassan} Sayed Fawad Hassan, Rachel A. Rosen, ``On Non-Linear
Actions for Massive Gravity'', {\hypersetup{urlcolor=vividviolet}\href{https://link.springer.com/article/10.1007\%2FJHEP07\%282011\%29009}%
{JHEP \textbf{07} (2011) 009}}, \href{https://arxiv.org/abs/1103.6055}{%
[arXiv:1103.6055 [hep-th]]}.

\bibitem{HassanI} Sayed Fawad Hassan, Rachel A. Rosen, ``Resolving the Ghost
Problem in Nonlinear Massive Gravity'', {\hypersetup{urlcolor=vividviolet}
\href{https://journals.aps.org/prl/abstract/10.1103/PhysRevLett.108.041101}{%
Phys. Rev. Lett. \textbf{108} (2012) 041101}}, \href{https://arxiv.org/abs/1106.3344}%
{[arXiv:1106.3344 [hep-th]]}.

\bibitem{HassanII} Sayed Fawad Hassan, Rachel A. Rosen, Angnis Schmidt-May,
``Ghost-free Massive Gravity with a General Reference Metric'', {%
\hypersetup{urlcolor=vividviolet}\href{https://link.springer.com/article/10.1007\%2FJHEP02\%282012\%29026}%
{JHEP \textbf{02} (2012) 026}}, \href{https://arxiv.org/abs/1109.3230}{%
[arXiv:1109.3230 [hep-th]]}.

\bibitem{1109.3515} S. F. Hassan, Rachel A. Rosen, ``Bimetric Gravity from
Ghost-free Massive Gravity'', {\hypersetup{urlcolor=vividviolet}\href{https://link.springer.com/article/10.1007\%2FJHEP02\%282012\%29126}%
{JHEP \textbf{02} (2012) 126}}, \href{https://arxiv.org/abs/1109.3515}{%
[arXiv:1109.3515 [hep-th]]}.

\bibitem{Vegh} David Vegh, ``Holography Without Translational Symmetry'',
\href{https://arxiv.org/abs/1301.0537}{[arXiv:1301.0537]}.

\bibitem{1308.4970} Mike Blake, David Tong, ``Universal Resistivity from
Holographic Massive Gravity'', {\hypersetup{urlcolor=vividviolet}\href{https://journals.aps.org/prd/abstract/10.1103/PhysRevD.88.106004}%
{Phys. Rev. D \textbf{88} (2013) 106004}}, \href{https://arxiv.org/abs/1308.4970}%
{[arXiv:1308.4970 [hep-th]]}.

\bibitem{1304.0484} Antonio De Felice, A. Emir Gumrukcuoglu, Chunshan Lin,
Shinji Mukohyama, ``On the Cosmology of Massive Gravity'', {%
\hypersetup{urlcolor=vividviolet}\href{http://iopscience.iop.org/article/10.1088/0264-9381/30/18/184004/meta}%
{Class. Quant. Grav. \textbf{30} (2013) 184004}}, \href{https://arxiv.org/abs/1304.0484}%
{[arXiv:1304.0484 [hep-th]]}.

\bibitem{1306.5457} Stanley Deser, Keisuke Izumi, Yen Chin Ong, Andrew
Waldron, ``Massive Gravity Acausality Redux'', {%
\hypersetup{urlcolor=vividviolet}\href{https://www.sciencedirect.com/science/article/pii/S0370269313007181?via\%3Dihub}%
{Phys. Lett. B \textbf{726} (2013) 544}}, \href{https://arxiv.org/abs/1306.5457}%
{[arXiv:1306.5457 [hep-th]]}.

\bibitem{1408.0561} Stanley Deser, McCullen Sandora, Andrew Waldron, George
Zahariade, ``Covariant Constraints for Generic Massive Gravity and Analysis
of Its Characteristics'', {\hypersetup{urlcolor=vividviolet}\href{https://journals.aps.org/prd/abstract/10.1103/PhysRevD.90.104043}%
{Phys. Rev. D 90 (2014) 104043}}, \href{https://arxiv.org/abs/1408.0561}{%
[arXiv:1408.0561 [hep-th]]}.

\bibitem{1410.2289} Stanley Deser, Keisuke Izumi, Yen Chin Ong, Andrew
Waldron, ``Problems of Massive Gravities'', {%
\hypersetup{urlcolor=vividviolet}\href{https://www.worldscientific.com/doi/abs/10.1142/S0217732315400064}%
{Mod. Phys. Lett. A30 (2015) 1540006}}, \href{https://arxiv.org/abs/1410.2289}%
{[arXiv:1410.2289 [hep-th]]}.

\bibitem{1504.02919} Stanley Deser, Andrew Waldron, George Zahariade,
``Propagation Peculiarities of Mean Field Massive Gravity'', {%
\hypersetup{urlcolor=vividviolet}\href{https://www.sciencedirect.com/science/article/pii/S0370269315005651?via\%3Dihub}%
{Phys. Lett. B \textbf{749} (2015) 144}}, \href{https://arxiv.org/abs/1504.02919}%
{[arXiv:1504.02919 [hep-th]]}.

\bibitem{1505.03518} Pavel Motloch, Wayne Hu, Austin Joyce, Hayato
Motohashi, ``Self-Accelerating Massive Gravity: Superluminality, Cauchy
Surfaces and Strong Coupling'', {\hypersetup{urlcolor=vividviolet}\href{https://arxiv.org/abs/1505.03518}%
{Phys. Rev. D \textbf{92} (2015) 044024}}, \href{https://arxiv.org/abs/1505.03518}%
{[arXiv:1505.03518 [hep-th]]}.

\bibitem{1706.07806} Sayed Fawad Hassan, Mikica Kocic, ``On the Local
Structure of Spacetime in Ghost-Free Bimetric Theory and Massive Gravity'', {%
\hypersetup{urlcolor=vividviolet}\href{https://link.springer.com/article/10.1007\%2FJHEP05\%282018\%29099}%
{JHEP \textbf{05} (2018) 099}}, \href{https://arxiv.org/abs/1706.07806}{%
[arXiv:1706.07806 [hep-th]]}.

\bibitem{1701.01039} Seyed Hossein Hendi, Gholam Hossein Bordbar, Behzad
Eslam Panah, Shahram Panahiyan, ``Neutron Stars Structure in the Context of
Massive Gravity'', {\hypersetup{urlcolor=vividviolet}\href{http://iopscience.iop.org/article/10.1088/1475-7516/2017/07/004/meta}%
{JCAP \textbf{07} (2017) 004}}, \href{https://arxiv.org/abs/1701.01039}{
[arXiv:1701.01039 [gr-qc]]}.

\bibitem{1805.10650} Behzad Eslam Panah, Helei Liu, `` White Dwarfs in de
Rham-Gabadadze-Tolley Like Massive Gravity'', {%
\hypersetup{urlcolor=vividviolet}\href{https://journals.aps.org/prd/abstract/10.1103/PhysRevD.99.104074}%
{Phys. Rev. D \textbf{99} (2019) 104074}}, \href{https://arxiv.org/abs/1805.10650}%
{[arXiv:1805.10650 [gr-qc]]}.

\bibitem{1808.06623} Koushik Dutta, Ruchika, Anirban Roy, Anjan A. Sen, M.M.
Sheikh-Jabbari, ``Beyond $\Lambda$CDM with Low and High Redshift Data:
Implications for Dark Energy'', \href{https://arxiv.org/abs/1808.06623}{%
[arXiv:1808.06623 [astro-ph.CO]]}.

\bibitem{1105.3735} Kurt Hinterbichler, ``Theoretical Aspects of Massive
Gravity'', {\hypersetup{urlcolor=vividviolet}\href{https://journals.aps.org/rmp/abstract/10.1103/RevModPhys.84.671}%
{Rev. Mod. Phys. \textbf{84} (2012) 671}},\href{https://arxiv.org/abs/1105.3735}%
{[ arXiv:1105.3735 [hep-th]]}.

\bibitem{Cai} Rong-Gen Cai, Ya-Peng Hu, Qi-Yuan Pan, Yun-Long Zhang,
``Thermodynamics of Black Holes in Massive Gravity'', {%
\hypersetup{urlcolor=vividviolet}\href{https://journals.aps.org/prd/abstract/10.1103/PhysRevD.91.024032}%
{Phys. Rev. D \textbf{91} (2015) 024032}}, \href{https://arxiv.org/abs/1409.2369}%
{[arXiv:1409.2369 [hep-th]]}.

\bibitem{Ghosh} Sushant G. Ghosh, Lunchakorn Tannukij, Pitayuth Wongjun,``A
Class of Black Holes in dRGT Massive Gravity and Their Thermodynamical
Properties'', {\hypersetup{urlcolor=vividviolet}\href{https://link.springer.com/article/10.1140\%2Fepjc\%2Fs10052-016-3943-x}%
{Eur. Phys. J. C \textbf{76} (2016) 119}}, \href{https://arxiv.org/abs/1506.07119}%
{[arXiv:1506.07119 [gr-qc]]}.

\bibitem{HendiEP1} Seyed Hossein Hendi, Behzad Eslam Panah, Shahram
Panahiyan, ``Einstein-Born-Infeld-Massive Gravity: AdS-Black Hole Solutions
and Their Thermodynamical Properties'', {\hypersetup{urlcolor=vividviolet}
\href{https://link.springer.com/article/10.1007\%2FJHEP11\%282015\%29157}{%
JHEP \textbf{11} (2015) 157}}, \href{https://arxiv.org/abs/1508.01311}{%
[arXiv:1508.01311 [hep-th]]}.

\bibitem{resultVII} Ping Li, Xin-zhou Li, Ping Xi, \textquotedblleft Black
Hole Solutions in de Rham-Gabadadze-Tolley Massive Gravity\textquotedblright
, {\hypersetup{urlcolor=vividviolet}\href{https://journals.aps.org/prd/abstract/10.1103/PhysRevD.93.064040}%
{Phys. Rev. D \textbf{93} (2016) 064040}}, \href{https://arxiv.org/abs/1603.06039}%
{[arXiv:1603.06039 [gr-qc]]}.

\bibitem{1510.03204} Hongsheng Zhang, Xin-Zhou Li, \textquotedblleft Ghost
Free Massive Gravity with Singular Reference Metrics\textquotedblright , {%
\hypersetup{urlcolor=vividviolet} \href{https://journals.aps.org/prd/abstract/10.1103/PhysRevD.93.124039}%
{Phys. Rev. D \textbf{93} (2016) 124039}}, \href{https://arxiv.org/abs/1510.03204}%
{[arXiv:1510.03204 [gr-qc]]}.

%\bibitem{Vainshtein1972} Arkady. I. Vainshtein, \textquotedblleft To the
%problem of nonvanishing gravitation mass\textquotedblright , {%
%\hypersetup{urlcolor=vividviolet}\href{https://www.sciencedirect.com/science/article/abs/pii/0370269372901475?via%3Dihub}%
%{Phys. Lett. B \textbf{39} (1972) 393}}.

%\bibitem{dRham2014} C. dRham, \textquotedblleft Black Hole Solutions in de
%Rham-Gabadadze-Tolley Massive Gravity\textquotedblright , {%
%\hypersetup{urlcolor=vividviolet}\href{https://link.springer.com/article/10.12942/lrr-2014-7}%
%{Living. Rev. Relativity \textbf{17} (2014) 7}}, \href{https://arxiv.org/abs/1401.4173}%
%{[arXiv:1401.4173 [hep-th]]}.

\bibitem{Ong2016} Yen Chin Ong, \textquotedblleft Hawking evaporation time
scale of topological black holes in anti-de Sitter
spacetime\textquotedblright , {\hypersetup{urlcolor=vividviolet}\href{https://www.sciencedirect.com/science/article/pii/S0550321316000067}%
{Nucl. Phys. B \textbf{903} (2016) 387}}, \href{https://arxiv.org/abs/1507.07845}%
{[arXiv:1507.07845 [gr-qc]]}.

\bibitem{Yao2019} Yuan Yao, Meng-Shi Hou, Yen Chin Ong, \textquotedblleft A
complementary third law for black hole thermodynamics\textquotedblright , {%
\hypersetup{urlcolor=vividviolet}} {\href{https://link.springer.com/article/10.1140/epjc/s10052-019-7003-1}%
{Eur. Phys. J. C 7\textbf{9} (2019) 513}}, \href{https://arxiv.org/abs/1812.03136}%
{[arXiv:1812.03136 [gr-qc]]}.

\bibitem{0106080} Ronald J. Adler, Pisin Chen, David I. Santiago,
\textquotedblleft The Generalized Uncertainty Principle and Black Hole
Remnants\textquotedblright , {\hypersetup{urlcolor=vividviolet}\href{https://link.springer.com/article/10.1023\%2FA\%3A1015281430411}%
{Gen. Rel. Grav. \textbf{33} (2001) 2101}}, \href{https://arxiv.org/abs/gr-qc/0106080}%
{[arXiv:gr-qc/0106080]}.

\bibitem{0912.2253} Petr Jizba, Hagen Kleinert, Fabio Scardigli,
``Uncertainty Relation on World Crystal and its Applications to Micro Black
Holes'', {\hypersetup{urlcolor=vividviolet}\href{https://journals.aps.org/prd/abstract/10.1103/PhysRevD.81.084030}%
{Phys. Rev. D \textbf{81} (2010) 084030}}, \href{https://arxiv.org/abs/0912.2253}%
{[arXiv:0912.2253 [hep-th]]}.

\bibitem{1806.03691} Yen Chin Ong, ``Zero Mass Remnant as an Asymptotic
State of Hawking Evaporation'', \href{https://arxiv.org/abs/1806.03691}{%
[arXiv:1806.03691 [gr-qc]]}.

\bibitem{0205106} Pisin Chen, Ronald J. Adler, ``Black Hole Remnants and
Dark Matter'', {\hypersetup{urlcolor=vividviolet}\href{https://www.sciencedirect.com/science/article/pii/S0920563203020887?via\%3Dihub}%
{Nucl. Phys. Proc. Suppl. \textbf{124} (2003) 103}}, \href{https://arxiv.org/abs/gr-qc/0205106}%
{[arXiv:gr-qc/0205106]}.

\bibitem{1207.3123} Ahmed Almheiri, Donald Marolf, Joseph Polchinski, James
Sully, ``Black Holes: Complementarity or Firewalls?'', {%
\hypersetup{urlcolor=vividviolet}\href{https://arxiv.org/abs/1207.3123}{JHEP
\textbf{02} (2013) 062}}, \href{https://arxiv.org/abs/1207.3123}{%
[arXiv:1207.3123 [hep-th]]}.

\bibitem{ACN} Yakir Aharonov, Aharon Casher, Shmuel Nussinov, ``The
Unitarity Puzzle and Planck Mass Stable Particles'', {%
\hypersetup{urlcolor=vividviolet}\href{https://www.sciencedirect.com/science/article/pii/0370269387913207}%
{Phys. Lett. B \textbf{191} (1987) 51}}.

\bibitem{0403104} Brett McInnes, ``Quintessential Maldacena-Maoz
Cosmologies'', {\hypersetup{urlcolor=vividviolet}\href{http://iopscience.iop.org/article/10.1088/1126-6708/2004/04/036/meta}%
{JHEP \textbf{04} (2004) 036}}, {\href{https://arxiv.org/abs/hep-th/0403104}{%
[arXiv:hep-th/0403104]}}.

\bibitem{1806.08362} Georges Obied, Hirosi Ooguri, Lev Spodyneiko, Cumrun
Vafa, ``De Sitter Space and the Swampland'', \href{https://arxiv.org/abs/1806.08362}%
{[arXiv:1806.08362 [hep-th]]}.

\bibitem{1808.09440} Yashar Akrami, Renata Kallosh, Andrei Linde, Valeri Vardanyan, ``The Landscape, the Swampland and the Era of Precision Cosmology'', 
{\hypersetup{urlcolor=vividviolet}\href{https://onlinelibrary.wiley.com/doi/full/10.1002/prop.201800075}{Fortsch. Phys. \textbf{67} (2019) 1800075}},
\href{https://arxiv.org/abs/1808.09440}{[arXiv:1808.09440 [hep-th]]}.

\bibitem{1811.05434} Federico Tosone, Balakrishna S. Haridasu, Vladimir V.
Lukovi\'c, Nicola Vittorio, ``Constraints on Field Flows of Quintessence
Dark Energy'', {\hypersetup{urlcolor=vividviolet}\href{https://journals.aps.org/prd/abstract/10.1103/PhysRevD.99.043503}%
{Phys. Rev. D 99 (2019) 043503}}, \href{https://arxiv.org/abs/1811.05434}{%
[arXiv:1811.05434 [astro-ph.CO]]}.

\bibitem{1808.08967} Michele Cicoli, Senarath de Alwis, Anshuman Maharana,
Francesco Muia, Fernando Quevedo, ``De Sitter vs Quintessence in String
Theory'', {\hypersetup{urlcolor=vividviolet}\href{https://onlinelibrary.wiley.com/doi/full/10.1002/prop.201800079}%
{Fortsch. Phys. \textbf{67} (2019) 1800079}}, \href{https://arxiv.org/abs/1808.08967}%
{[arXiv:1808.08967 [hep-th]]}.

\bibitem{1812.03184} Mark P. Hertzberg, McCullen Sandora, Mark Trodden,
``Quantum Fine-Tuning in Stringy Quintessence Models'', {%
\hypersetup{urlcolor=vividviolet}\href{https://www.sciencedirect.com/science/article/pii/S0370269319305921?via\%3Dihub}%
{Phys. Lett. B \textbf{797} (2019) 134878}}, \href{https://arxiv.org/abs/1812.03184}%
{[arXiv:1812.03184 [hep-th]]}.

\bibitem{1902.11014v2} Suddhasattwa Brahma, Md. Wali Hossain, ``Dark Energy
Beyond Quintessence: Constraints From the Swampland'', {%
\hypersetup{urlcolor=vividviolet}\href{https://link.springer.com/article/10.1007\%2FJHEP06\%282019\%29070}%
{JHEP \textbf{06} (2019) 070}}, \href{https://arxiv.org/abs/1902.11014v2}{%
[arXiv:1902.11014 [hep-th]]}.

\bibitem{1610.01505} Seyed Hossein Hendi, Nematollah Riazi, Shahram
Panahiyan, ``Holographical Aspects of Dyonic Black Holes: Massive Gravity
Generalization'', {\hypersetup{urlcolor=vividviolet}\href{https://onlinelibrary.wiley.com/doi/abs/10.1002/andp.201700211}%
{Ann. Phys. (Berlin) \textbf{530} (2018) 1700211}}, \href{https://arxiv.org/abs/1610.01505}%
{[arXiv:1610.01505 [hep-th]]}.

\bibitem{1709.07503} Jun Zhang, Shuang-Yong Zhou, ``Can the Graviton Have a
Large Mass Near Black Holes?'', {\hypersetup{urlcolor=vividviolet}\href{https://journals.aps.org/prd/abstract/10.1103/PhysRevD.97.081501}%
{Phys. Rev. D \textbf{97} (2018) 081501}}, \href{https://arxiv.org/abs/1709.07503}%
{[arXiv:1709.07503 [gr-qc]]}.

\bibitem{9305007} Don N. Page, ``Average Entropy of a Subsystem'', {%
\hypersetup{urlcolor=vividviolet}\href{https://journals.aps.org/prl/abstract/10.1103/PhysRevLett.71.1291}%
{Phys. Rev. Lett. \textbf{71} (1993) 1291}}, \href{https://arxiv.org/abs/gr-qc/9305007}%
{[arXiv:gr-qc/9305007]}.

\bibitem{9306083} Don N. Page, ``Information in Black Hole Radiation'', {%
\hypersetup{urlcolor=vividviolet}\href{https://journals.aps.org/prl/abstract/10.1103/PhysRevLett.71.3743}%
{Phys. Rev. Lett. \textbf{71} (1993) 3743}}, \href{https://arxiv.org/abs/hep-th/9306083}%
{[arXiv:hep-th/9306083]}.

\bibitem{1301.4995} Don N. Page, ``Time Dependence of Hawking Radiation
Entropy'', {\hypersetup{urlcolor=vividviolet}\href{http://iopscience.iop.org/article/10.1088/1475-7516/2013/09/028/meta}%
{JCAP \textbf{09} (2013) 028}}, \href{https://arxiv.org/abs/1301.4995}{%
[arXiv:1301.4995 [hep-th]]}.

\bibitem{1506.03975} Finnian Gray, Sebastian Schuster, Alexander Van-Brunt,
Matt Visser, ``The Hawking Cascade from a Black Hole is Extremely Sparse'', {%
\hypersetup{urlcolor=vividviolet}\href{http://iopscience.iop.org/article/10.1088/0264-9381/33/11/115003/meta}%
{Class. Quant. Grav. \textbf{33} (2016) 115003}}, \href{https://arxiv.org/abs/1506.03975}%
{[arXiv:1506.03975 [gr-qc]]}.

\bibitem{1512.05809} Matt Visser, Finnian Gray, Sebastian Schuster,
Alexander Van-Brunt, ``Sparsity of the Hawking Flux'', Proceedings of the
MG14 Meeting on General Relativity (2017); pp. 1724-1729, \href{https://arxiv.org/abs/1512.05809}%
{[arXiv:1512.05809 [gr-qc]]}.

\bibitem{1503.07521} Yashar Akrami, S. F. Hassan, Frank K\"onnig, Angnis
Schmidt-May, Adam R. Solomon, ``Bimetric Gravity is Cosmologically Viable'',
{\hypersetup{urlcolor=vividviolet}\href{https://www.sciencedirect.com/science/article/pii/S0370269315004840?via\%3Dihub}%
{Phys. Lett. B \textbf{748} (2015) 37}}, \href{https://arxiv.org/abs/1503.07521}%
{[arXiv:1503.07521 [gr-qc]]}.

\bibitem{1802.07267} Sayed Fawad Hassan, Anders Lundkvist, ``Analysis of
Constraints and Their Algebra in Bimetric Theory'', \href{https://arxiv.org/abs/1802.07267}%
{[arXiv:1802.07267 [hep-th]]}.

\bibitem{MacGibbon} Jane H. MacGibbon, ``Can Planck-Mass Relics of
Evaporating Black Holes Close the Universe?'', {%
\hypersetup{urlcolor=vividviolet}\href{https://www.nature.com/articles/329308a0}%
{Nature \textbf{329} (1987) 308}}.

\bibitem{Barrow} John D. Barrow, Edmund J. Copeland, Andrew R. Liddle, ``The
Cosmology of Black Hole Relics'', {\hypersetup{urlcolor=vividviolet}\href{https://journals.aps.org/prd/abstract/10.1103/PhysRevD.46.645}%
{Phys. Rev. D \textbf{46} (1992) 645}}.

\bibitem{Carr} Bernard J. Carr, Jonathan H. Gilbert, James E. Lidsey,
``Black Hole Relics and Inflation: Limits on Blue Perturbation Spectra'', {%
\hypersetup{urlcolor=vividviolet}\href{https://journals.aps.org/prd/abstract/10.1103/PhysRevD.50.4853}%
{Phys. Rev. D \textbf{50} (1994) 4853}}, \href{https://arxiv.org/abs/astro-ph/9405027}%
{[arXiv:astro-ph/9405027]}.

\bibitem{0411158} Sergei L. Dubovsky, Peter G. Tinyakov, Igor I. Tkachev,
``Massive Graviton as a Testable Cold Dark Matter Candidate'', {%
\hypersetup{urlcolor=vividviolet}\href{https://journals.aps.org/prl/abstract/10.1103/PhysRevLett.94.181102}%
{Phys. Rev. Lett. \textbf{94} (2005) 181102}}, \href{https://arxiv.org/abs/hep-th/0411158}%
{[arXiv:hep-th/0411158]}.

\bibitem{1604.06704} Katsuki Aoki, Shinji Mukohyama, ``Massive Gravitons as
Dark Matter and Gravitational Waves'', {\hypersetup{urlcolor=vividviolet}
\href{https://journals.aps.org/prd/abstract/10.1103/PhysRevD.94.024001}{%
Phys. Rev. D \textbf{94} (2016) 024001}}, \href{https://arxiv.org/abs/1604.06704}%
{[arXiv:1604.06704 [hep-th]]}.

\bibitem{1604.08564} Eugeny Babichev, Luca Marzola, Martti Raidal, Angnis
Schmidt-May, Federico Urban, Hardi Veerm\"ae, Mikael von Strauss,
``Gravitational Origin of Dark Matter'', {\hypersetup{urlcolor=vividviolet}
\href{https://journals.aps.org/prd/abstract/10.1103/PhysRevD.94.084055}{%
Phys. Rev. D \textbf{94} (2016) 084055}}, \href{https://arxiv.org/abs/1604.08564}%
{[arXiv:1604.08564 [hep-ph]]}.

\bibitem{1607.03497} Eugeny Babichev, Luca Marzola, Martti Raidal, Angnis
Schmidt-May, Federico Urban, Hardi Veerm\"ae, Mikael von Strauss,
\textquotedblleft Heavy Spin-2 Dark Matter\textquotedblright , {%
\hypersetup{urlcolor=vividviolet}\href{http://iopscience.iop.org/article/10.1088/1475-7516/2016/09/016/meta}%
{JCAP \textbf{09} (2016) 016}}, \href{https://arxiv.org/abs/1607.03497}{%
[arXiv:1607.03497 [hep-th]]}.

\bibitem{1708.04253} Luca Marzola, Martti Raidal, Federico R. Urban,
``Oscillating Spin-2 Dark Matter'', {\hypersetup{urlcolor=vividviolet}\href{https://journals.aps.org/prd/abstract/10.1103/PhysRevD.97.024010}%
{Phys. Rev. D \textbf{97} (2018) 024010}}, \href{https://arxiv.org/abs/1708.04253}%
{[arXiv:1708.04253 [hep-ph]]}.
\end{thebibliography}
\end{document}